\documentclass[journal]{IEEEtran}
\usepackage{graphicx}

\begin{document}
%\title{Experimental Validation of K and L shell Radiative Transition Probabilities}
\title{Validation of fluorescence transition probability calculations}

\author{Maria Grazia Pia,
	Paolo Saracco
        and Manju Sudhakar% <-this % stops a space
\thanks{M. G. Pia and P. Saracco are with INFN Sezione di Genova, 
	Via Dodecaneso 33, 16146 Genova, Italy 
	(e-mail: MariaGrazia.Pia@ge.infn.it, 
	Paolo.Saracco@ge.infn.it)}% <-this % stops a space
\thanks{M. Sudhakar is with INFN Sezione di Genova, 
	Via Dodecaneso 33, 16146 Genova, Italy 
	and Department of Physics, University of Calicut, India
	(e-mail: Manju.Sudhakar@ge.infn.it); 
	she is on leave from ISRO Bangalore, India}% <-this % stops a space

\thanks{Manuscript received November16, 2009.}}
\maketitle
\pagestyle{empty}
\thispagestyle{empty}

\begin{abstract}
%\boldmath
A systematic and quantitative validation of the K and L shell X-ray transition
probability calculations according to different theoretical methods has 
been performed against experimental data. 
This study  is relevant to the optimization of data libraries
%based on theoretical calculations, that are 
used by software systems, 
namely Monte Carlo codes, dealing with X-ray fluorescence.
The results support the adoption of transition probabilities calculated 
according to the Hartree-Fock approach, which manifest better agreement 
with experimental measurements than calculations based on the Hartree-Slater
method.

%The aim of this work
%is to compare between two theoretical approaches for computation of the
%transition probabilities, one of which is being currently used in the Evaluated
%Atomic Data Libraries (EADL), and the other employs certain corrections
%pertinent to Fermi-Dirac statistics and exchange and overlap effects. This
%validation is essential as it will give an idea of which is the relatively
%better model, and can be used to update the values in EADL, thus providing the
%experimental community with a fairly accurate data base that will aid in the
%simulation of X-ray fluorescence.

\end{abstract}

\begin{IEEEkeywords}
X-ray fluorescence, PIXE, Monte Carlo.
\end{IEEEkeywords}

\IEEEpeerreviewmaketitle

% ----------------------------------------------------------------------------
\section{Introduction}

\IEEEPARstart{A}{nalysis} techniques using X-ray fluorescence are
non-destructive methods to determine the elemental composition 
of material samples in a variety of
applications, from planetary science to cultural heritage.

Software systems that deal with X-ray fluorescence, either for
elemental analysis or Monte Carlo simulation, require accurate
values of the physics parameters relevant for this process: the
cross sections for the occurrence of the primary process creating a
vacancy in the shell occupancy, the probability of radiative
transitions once a vacancy has been created, and the energy of the
emitted X-rays, which is determined by the binding energies of the
atomic levels involved in the transition.
These quantities usually derive from theoretical calculations, since
experimental measurements cannot practically cover the entire range of
physics conditions (target elements and incident particle characteristics)
required by general-purpose software systems.
The results of theoretical calculations are often tabulated in data libraries
to avoid time-consuming computations of complex analytical formulae
in software applications.

Calculations of radiative transition probabilities according to two different
approaches, based on the Hartree-Slater and Hartree-Fock methods,
are documented in the literature \cite{sco0,sco1,sco2,sco3}.
Tabulations deriving from calculations with the Hartree-Slater method
are collected in the Evaluated Atomic Data Library (EADL) \cite{eadl}, 
which is used by various Monte Carlo codes, 
including Geant4 \cite{g4nim,g4tns},
%like EGSnrc
%\cite{egsnrc}, Geant4 \cite{g4nim,g4tns}, MCNP \cite{mcnp5} and Penelope
%\cite{penelope}, 
for the simulation of X-ray fluorescence.
GUPIX \cite{gupix1,gupix2}, a specialized software system which is widely
used for elemental analysis with PIXE (Particle Induced X-ray Emission)
techniques, instead uses a database of K and L X-ray intensities based
on Hartree-Fock calculations; however, this code and its
databases are not freely available.

A systematic and
quantitative evaluation of the relative merits of the two theoretical
methods with respect to an extensive data sample is not available yet.
This issue has been addressed by the study documented in this paper.

%The Evaluated Atomic Data Library (EADL)\cite{eadl} has been used in different
%multi-particle Monte Carlo simulation systems, like PENELOPE\cite{pene},
%EGSnrc\cite{egs}, MCNP\cite{mcnp}, Geant4\cite{geant1}, \cite{geant2}, to
%simulate the relaxation of an ionized atom to the ground state, with the
%emission of X-rays (fluorescence) and/or electrons (Auger effect). 

%In this paper, the authors aim to conduct a systematic validation of the K and L 
%shell radiative transition probabilities in EADL with experimental data.

% ----------------------------------------------------------------------------
\section{Theoretical background}

If one considers two energy levels in an atom, a perturbation to the
system (like excitation or ionization) results in a superposition of
the wavefunctions of the two levels; this superposition manifests
itself as a probability amplitude or a charge cloud.
This charge cloud oscillates with a frequency that is equal to the
energy difference between the two states, causing the emission of
radiation.
If this disturbed system consists of only one electron, there is only
the interaction between the nucleus and the electron to consider, and
this can be described by a $1/r$ potential; for a many-electron system
the repulsive force between the electron in question and the other
electrons in the atom should also be included.
This repulsive force is assumed to act centrally, like the $1/r$ force
between the electron and the nuclues; combining these two, one can
define the central field.
The structure of this field is a function of the effective charge
Z$_{eff}$ of the screened nucleus and this screening, hence Z$_{eff}$ is 
%naturally
a function of the effective distance $r$ of the electron from the nucleus.
This field can be determined by what is called the ``self consistent
field'' method: an initial guess about the form of this field
is made, which is used in the time-dependent Schr\"odinger
equation to compute the wavefunctions;
these are then used to calculate the charge distribution, and finally
the potential set up by the charge distribution is determined.
If the initial guess and the computed value do not match, the
process is iterated.

This calculation was first made using the Hartree-Slater approach, where
the electrons are assumed to move independently, with their mutual
interaction accounted for by a mean field central potential; electrons,
moreover, are treated relativistically and the effect of retardation is
included \cite{sco2}. 
However, within this approach initial and final wave functions are
assumed to be identical, therefore missing some of the effects induced by the
the Fermi statistics. 
The restricted Hartree-Fock approach was an obvious
correction, giving a more accurate estimate of
matrix elements of the transition operator between different subshells 
\cite{sco2,sco3}:
the improvement comes essentially because there is room for a non vanishing
overlap integral between initial and final single particle wave functions,
which now are not assumed to be identical.

% ----------------------------------------------------------------------------
\section{Overview of the validation analysis}

The validation study involved the comparison of fluorescence
transition probabilities deriving from theoretical calculations against
experimental measurements.

The theoretical and experimental data relevant to this study
are available under various forms in the literature:
\begin{itemize}
\item radiative emission rates, i.e. rates of decays of vacancies in a given 
shell accompanied by the emission of X-rays,
\item ratios of radiative emission rates, where the numerator and
denominator in the ratios may concern an individual transition or a set of 
transitions,
\item probabilities of radiative transitions concerning individual shells,
normalized over both radiative and non-radiative transitions.
\end{itemize}
The various data references cover different sets of transitions.

The different types of data were converted into a consistent representation to
allow their comparison: transition probabilities over
a common subset of transitions, listed in Table \ref{tab_trans}.

The experimental data were extracted from the compilation of
references in \cite{salem}; they were subject to selection and
normalization procedures.

Theoretical values calculated according to the Hartree-Slater and
Hartree-Fock approaches were taken from \cite{sco0}-\cite{sco3}; 
the selected subset of theoretical values corresponds to
the transitions for which experimental data are reported in
\cite{salem}.

The radiative transition probabilities in EADL were processed
similarly to the other theoretical tabulations. 

For each transition, the theoretical and EADL transition probabilities as
functions of the atomic number Z were compared with the experimental references
using statistical methods to estimate their compatibility.
The $\chi^{2}$ \cite{bock} test was performed to compare the data for 
each element.
The null-hypothesis in the $\chi^{2}$ test assumed that the
experimental data and those based on theoretical calculations derive
from the same parent population; a 0.05 significance level was set to
define the critical region of rejection of the null hypothesis.

Contingency tables were exploited to analyze the data resulting from the
outcome of the $\chi^{2}$ test for each category of theoretical data.
They were built
based on the number of transitions that pass or fail the $\chi^{2}$
test, i.e. for which the p-value resulting from the test is greater or 
smaller than 0.05.
In the analysis of the contingency tables the null hypothesis assumed the
categories under evaluation to be equivalent regarding their accuracy to
reproduce the experimental data.
Contingency tables were analyzed with Fisher's exact test
\cite{fisher}; as a cross-check, a $\chi^{2}$ test was also 
performed on the contingency tables, applying Yates' correction
\cite{yates} to account for the small number of entries in the tables.

% ----------------------------------------------------------------------------
\section{Experimental data}
\label{sec_selamdata}

An extensive compilation of experimental emission rates ratios
for the K and L shell transitions is documented in \cite{salem}.
To date it is still the most complete
source of K and L shell experimental transition probability
ratios; its relevance is confirmed by the fact that a recent database
for X-ray spectroscopy \cite{elam}, available from the NIST (National
Institute of Standards) \cite{elamnist}, is based on it for what
concerns K and L radiative transition probabilities.
Later measurements \cite{dost,raulo} have been found consistent with
the content of \cite{salem}.

\begin{table}[!t]
\caption{Radiative transition probabilities examined in this study}
\label{tab_trans}
\centering
\begin{tabular}{|l|c|}
\hline
{\bf Transitions measured against a reference}	& {\bf Reference} 	      \\
\hline		                      
K-L$_{2}$, K-M$_{2}$, K-M$_{3}$, K-M$_{4,5}$, K-N$_{2,3}$, K-N$_{4,5}$ & K-L$_{3}$         \\
L$_{1}$-M$_{2}$, L$_{1}$-N$_{2}$, L$_{1}$-N$_{3}$   			& L$_{1}$-M$_{3}$  \\
L$_{2}$-M$_{1}$, L$_{2}$-N$_{4}$, L$_{2}$-O$_{4}$  			& L$_{2}$-M$_{4}$  \\
L$_{3}$-M$_{1}$, L$_{3}$-M$_{4}$, L$_{3}$-N$_{1}$, L$_{3}$-N$_{4,5}$, L$_{3}$-O$_{4,5}$ & L$_{3}$-M$_{5}$ \\     
\hline			  	        
\end{tabular}
\end{table}

\subsection{Experimental sample}

The original experimental measurements compiled in \cite{salem}
were used for the validation of the theory. 

Since \cite{salem} reports tabulations of fits to the data, but only
bibliographical references and graphics of the original data, 
the experimental measurements and their uncertainties were retrieved from 
the original
references, whenever they reported numerical values, or digitized from
the published figures in
cases where only graphical representations were available.
The DigitizeIt \cite{digi} software was used for this purpose.
%The compilation reports the original experimental data only in
%graphical form and the corresponding bibliographical references.  

The uncertainties introduced by the digitization process were estimated
by comparing the published numerical data, when available in the original
references, to the corresponding digitized values.
The average difference between published and digitized values was 
verified to be smaller than 2\%, with the exception of the $L_{2}M_{1}$
transition, where 5\% differences where observed.
A further verification was performed on a selected data sample
by comparing the values digitized by two software systems,
DigitizeIt  and Engauge \cite{engauge}; the relative difference
was smaller than 2\%.

The experimental uncertainties for the emission rates vary from a few percent
for the K-L$_{2}$ transition to approximately 25\% for the
L$_{2}$-O$_{4}$ transition.
The uncertainties associated with some of the experimental data
are not specified in the original references,
nor in \cite{salem};
they were assumed in this study to be consistent with
the average errors reported in other publications for the same kind of
measurements and similar experimental conditions.
In a few cases where such an inference was not possible due to the
lack of comparable measurements,
the data points deprived of any error estimate were not considered in
the validation process.

Further evaluations were performed on the experimental collection to
select a data sample suitable to be used as a reference for the
validation of the theory.
Those points not included in the data fits of \cite{salem} were discarded 
consistently with the arguments discussed in that reference.
Multiple experimental data for the same element were combined;
data points identified as outliers were discarded.
Experimental data series looking
largely inconsistent with the data collected by other experiments,
were discarded too, as presumably affected by systematic errors.

%A similar criterion was also applied with respect to data points corresponding 
%to neighbouring atomic numbers: in this case the 3$\sigma$ criterion was
%applied with respect to the predicted value resulting from the interpolation
%of the neighboring values. 

\subsection{Determination of transition probabilities}
\label{sec_fit}

The data derived from this selection were subject to a preliminary
treatment, to determine individual transition probabilities from the
tabulated emission rate ratios.
Since the experimental values reported in \cite{salem} are ratios
relative to the strongest line in the series, the emission rate of
the strongest line of each series was assumed to be one.
Then the emission rates of each transition were normalized with
respect to the sum of the emission rates of all the transitions in the
series.

A method was devised to perform an indirect evaluation of experimental
compatibility also for those transitions associated with the strongest
line in each series, which have been taken as a reference in the
probability ratios reported in \cite{salem}.
The experimental reference probabilities for these transitions were
calculated as the complement to unit total probability, taking
into account the values associated with the other measured transitions.

The least square fits to the data tabulated in \cite{salem} and 
further interpolated in \cite{elam} were retained
for most transitions; in a few cases ($L_{1}N_{2}$, $L_{2}N_{4}$ and 
$L_{2}O_{4}$ transitions) improved fits were found to better
describe the data.

% ---------------------------------------------------------------------------- 
\section{Theoretical data}

Emission rates deriving from Hartree-Slater and
Hartree-Fock calculations, and EADL tabulations were subject to 
preliminary processing to assemble theoretical samples suitable
to validation against the experimental data derived from \cite{salem}.
The procedures adopted to retrieve theoretical data sets 
for the transitions listed in Table \ref{tab_trans}, out of the
theoretical tabulations available in the literature, are described in detail
in \cite{trans_tns}.

\subsection{Emission rates based on the Hartree-Slater method}
\label{sec_hs}

K and L X-ray emission rates, calculated by Scofield using the
Hartree-Slater approach for elements with atomic number from 5 to 104,
are tabulated in \cite{sco1}.

Only those transitions that are listed in Table \ref{tab_trans} were
selected for the validation process; the data were subject to
normalization to obtain transition probabilities for each element
relative to the subset of transitions under study (i.e. to
each row in Table  \ref{tab_trans}).

\subsection{Emission rates based on the Hartree-Fock method}
\label{sec_hf}
	
The K shell and L shell X-ray emission rates calculated by Scofield
using the Hartree-Fock approach are tabulated in \cite{sco2} and \cite{sco3}.
For the K shell, the emission rate ratios with respect to the
strongest line in the series are listed for 50 elements with atomic
number between 10 and 98;
a limited number of emission rate ratios are reported.

The available data were transformed into transition probabilities;
for each element the probabilities were normalized to 1 over each row
of Table \ref{tab_trans}.

For the L-shell emission rates, the tabulations in \cite{sco3} are 
listed only for 21 elements in the range 18$\leq$Z$\leq$94.
These have been computed from the Hartree-Fock based emission rates and
then fitted with polynomials as a function of Z in \cite{puri};  
the coefficients of these polynomials are reported in this reference for the
different ranges of Z over which they are valid, for each transition.
Using these coefficients, one can form equations to
compute the intensities relative to the strongest line in the series; the
absolute value of the strongest line of the series is also provided in
\cite{puri}, using which the individual transition probability for each
transition can be computed for all Z.

\subsection{Transition probabilities in EADL}
\label{sec_eadl}

EADL includes binding energies of electrons for all subshells, the transition
probabilities between subshells for emission of fluorescence photons and Auger
electrons, and the energy of these emitted particles, for Z from 6 to 100.
The transition probabilities are for all filled subshells in a neutral atom;
it is assumed that the atomic relaxation process following the creation of
an initial vacancy is independent of the ionizing radiation.

%The K and L shell emission rates have been calculated by Scofield, using both
%the Hartree-Slater potential \cite{sco0}\cite{sco1} and the Hartree-Fock
%potential \cite{sco2}\cite{sco3}.
According to \cite{eadl}, the EADL radiative transition probabilities have been
derived from Scofield's Hartree-Slater calculations \cite{sco0,sco1}.
Given the unclear documentation of the source of EADL tabulations, the
validation process was meant not only to estimate the accuracy of this data
library, but also to ascertain its content with respect to the
published theoretical references.
	
The sum of the radiative and non-radiative transition probabilities
listed in EADL
for a shell (or subshell) adds to one for a particular element.  

The transitions listed in EADL are extensive compared to those in \cite{salem};
only those transitions in EADL that are common with those in \cite{salem} were
considered in the validation process, as listed in Table \ref{tab_trans}.

% ----------------------------------------------------------------------------
\section{Results}
\label{results}

\begin{figure}[!t]
\centering
\includegraphics[width=2.5in,angle=-90]{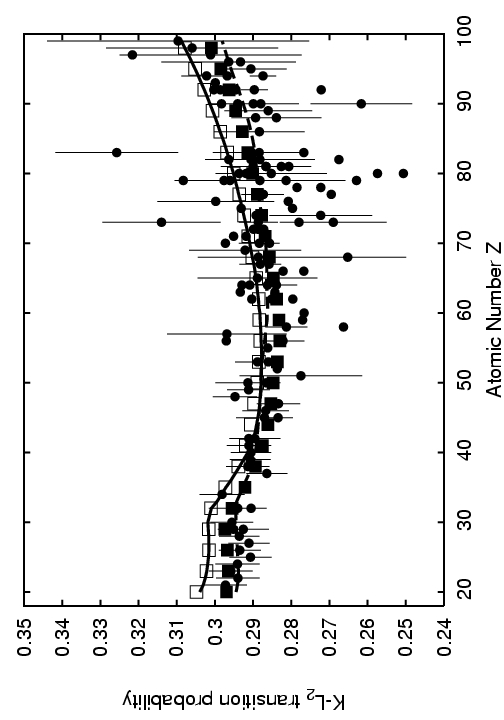}
\caption{K-L$_{2}$ transition probability versus Z:
theoretical calculations based on the Hartree-Slater \cite{sco1}
(white squares) and the Hartree-Fock \cite{sco2} (black squares)
potentials, EADL \cite{eadl} tabulations (solid line), experimental
data (black circles) and fit to them as in \cite{salem} (dashed line).}
\label{fig-kl2}
\end{figure}

\begin{figure}[!t]
\centering
\includegraphics[width=2.5in,angle=-90]{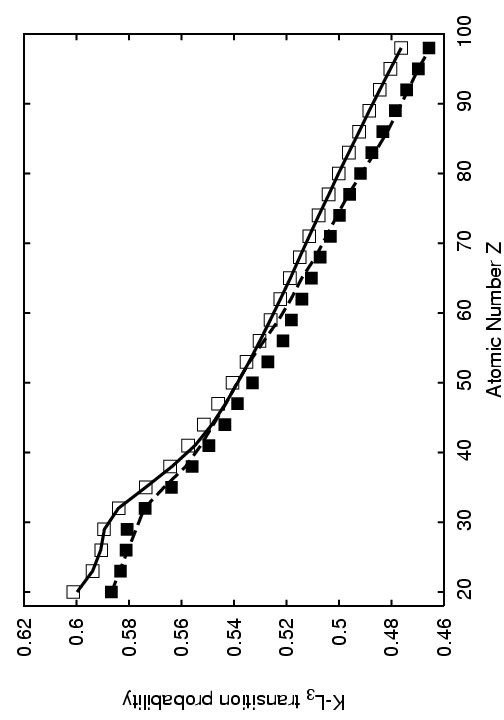}
\caption{K-L$_{3}$ transition probability versus Z:
theoretical calculations based on the Hartree-Slater \cite{sco1}
(white squares) and the Hartree-Fock \cite{sco2} (black squares)
potentials, EADL \cite{eadl} tabulations (solid line) and fit to experimental
data as in \cite{salem} (dashed line).}
\label{fig-kl3}
\end{figure}

\begin{figure}[!t]
\centering
\includegraphics[width=2.5in,angle=-90]{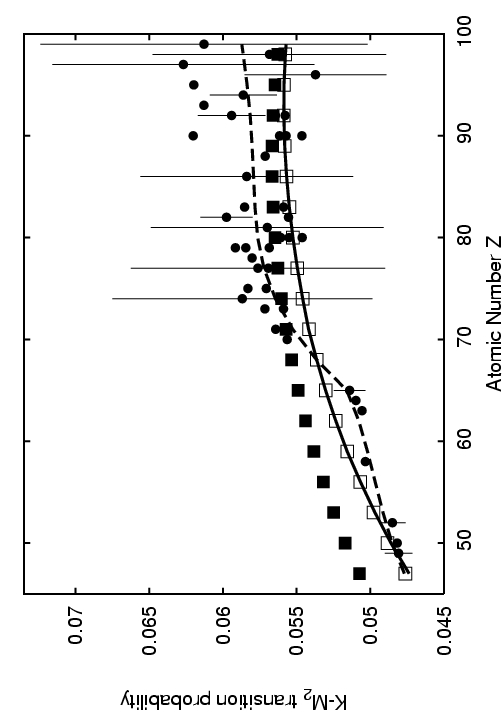}
\caption{K-M$_{2}$ transition probability versus Z:
theoretical calculations based on the Hartree-Slater \cite{sco1}
(white squares) and the Hartree-Fock \cite{sco2} (black squares)
potentials, EADL \cite{eadl} tabulations (solid line), experimental
data (black circles) and fit to them as in \cite{salem} (dashed line).}
\label{fig-km2}
\end{figure}

\begin{figure}[!t]
\centering
\includegraphics[width=2.5in,angle=-90]{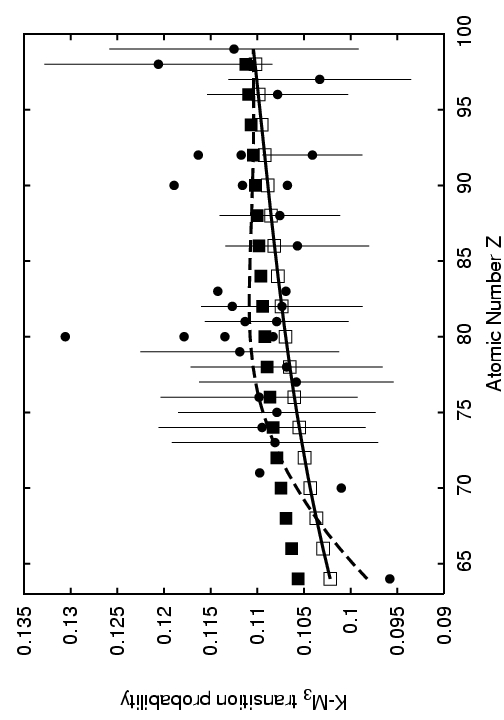}
\caption{K-M$_{3}$ transition probability versus Z:
theoretical calculations based on the Hartree-Slater \cite{sco1}
(white squares) and the Hartree-Fock \cite{sco2} (black squares)
potentials, EADL \cite{eadl} tabulations (solid line), experimental
data (black circles) and fit to them as in \cite{salem} (dashed line).}
\label{fig-km3}
\end{figure}

\begin{figure}[!t]
\centering
\includegraphics[width=2.5in,angle=-90]{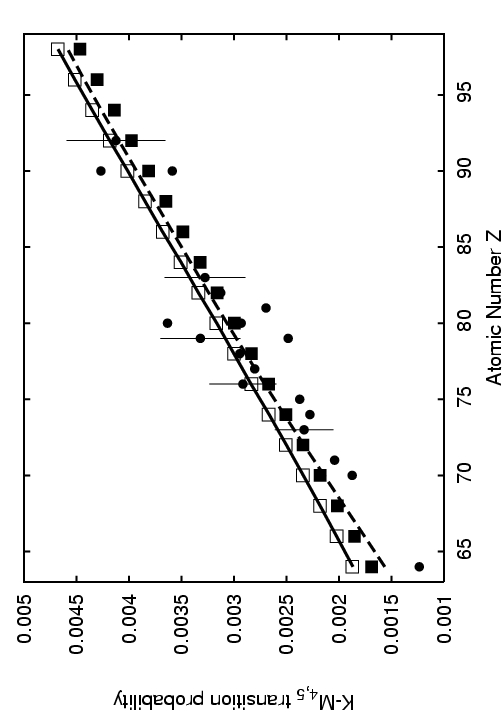}
\caption{K-M$_{4,5}$ transition probability versus Z:
theoretical calculations based on the Hartree-Slater \cite{sco1}
(white squares) and the Hartree-Fock \cite{sco2} (black squares)
potentials, EADL \cite{eadl} tabulations (solid line), experimental
data (black circles) and fit to them as in \cite{salem} (dashed line).}
\label{fig-km45}
\end{figure}

\begin{figure}[!t]
\centering
\includegraphics[width=2.5in,angle=-90]{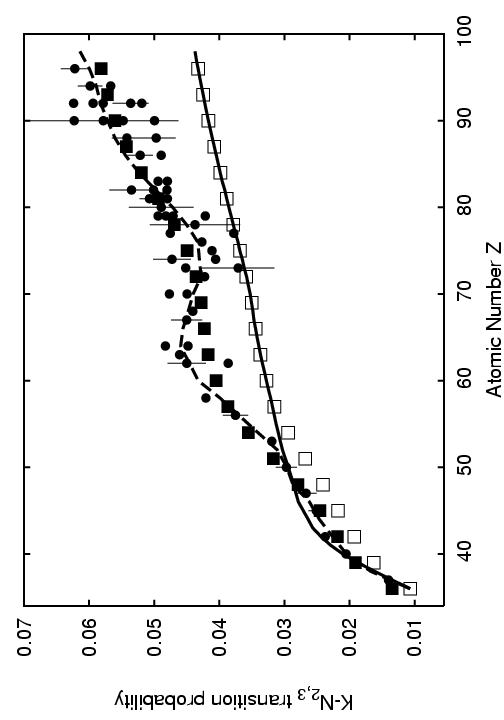}
\caption{K-N$_{2,3}$ transition probability versus Z:
theoretical calculations based on the Hartree-Slater \cite{sco1}
(white squares) and the Hartree-Fock \cite{sco2} (black squares)
potentials, EADL \cite{eadl} tabulations (solid line), experimental
data (black circles) and fit to them as in \cite{salem} (dashed line).}
\label{fig-kn23}
\end{figure}

\begin{figure}[!t]
\centering
\includegraphics[width=2.5in,angle=-90]{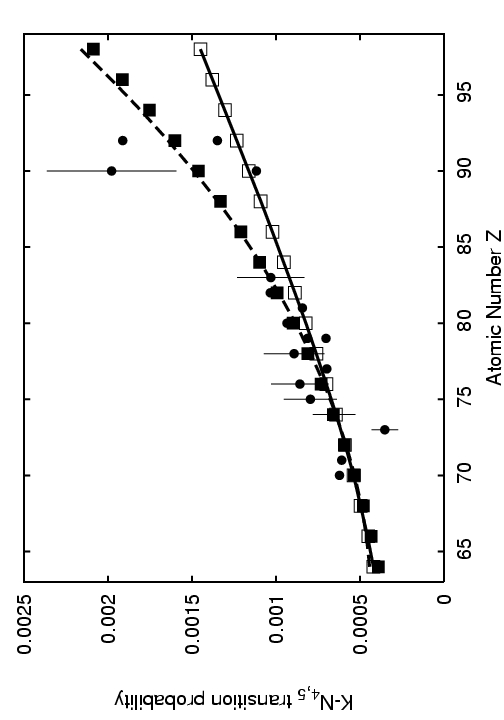}
\caption{K-N$_{4,5}$ transition probability versus Z:
theoretical calculations based on the Hartree-Slater \cite{sco1}
(white squares) and the Hartree-Fock \cite{sco2} (black squares)
potentials, EADL \cite{eadl} tabulations (solid line), experimental
data (black circles) and fit to them as in \cite{salem} (dashed line).}
\label{fig-kn45}
\end{figure}

\begin{figure}[!t]
\centering
\includegraphics[width=2.5in,angle=-90]{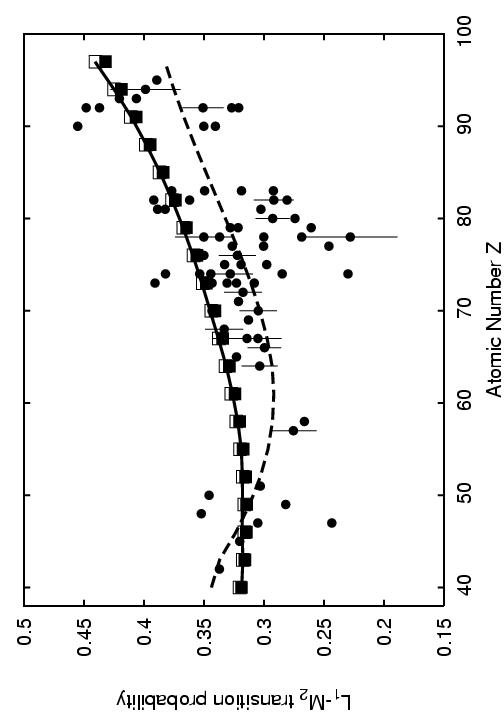}
\caption{L$_{1}$-M$_{2}$ transition probability versus Z:
theoretical calculations based on the Hartree-Slater \cite{sco1}
(white squares) and the Hartree-Fock \cite{sco2} (black squares)
potentials, EADL \cite{eadl} tabulations (solid line), experimental
data (black circles) and fit to them as in \cite{salem} (dashed line).}
\label{fig-l1m2}
\end{figure}

\begin{figure}[!t]
\centering
\includegraphics[width=2.5in,angle=-90]{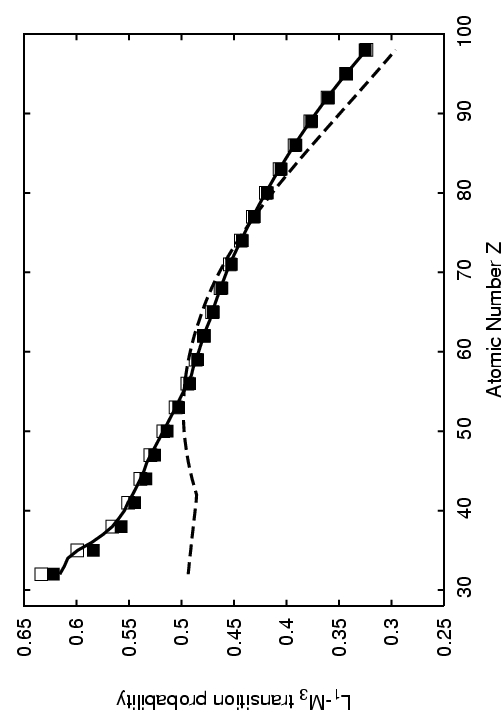}
\caption{L$_{1}$-M$_{3}$ transition probability versus Z:
theoretical calculations based on the Hartree-Slater \cite{sco1}
(white squares) and the Hartree-Fock \cite{sco2} (black squares)
potentials, EADL \cite{eadl} tabulations (solid line) and fit to experimental
data as in \cite{salem} (dashed line).}
\label{fig-l1m3}
\end{figure}

\begin{figure}[!t]
\centering
\includegraphics[width=2.5in,angle=-90]{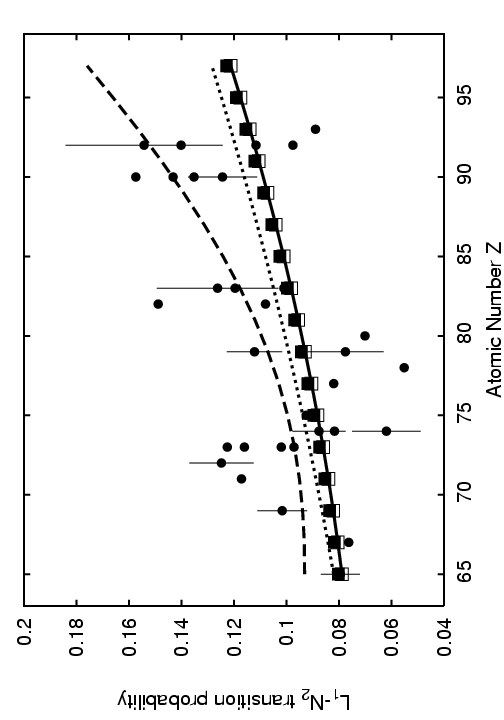}
\caption{L$_{1}$-N$_{2}$ transition probability versus Z:
theoretical calculations based on the Hartree-Slater \cite{sco1}
(white squares) and the Hartree-Fock \cite{sco2} (black squares)
potentials, EADL \cite{eadl} tabulations (solid line), experimental
data (black circles), fit to them as in \cite{salem} (dashed line), and
improved fit (dotted line).}
\label{fig-l1n2}
\end{figure}

\begin{figure}[!t]
\centering
\includegraphics[width=2.5in,angle=-90]{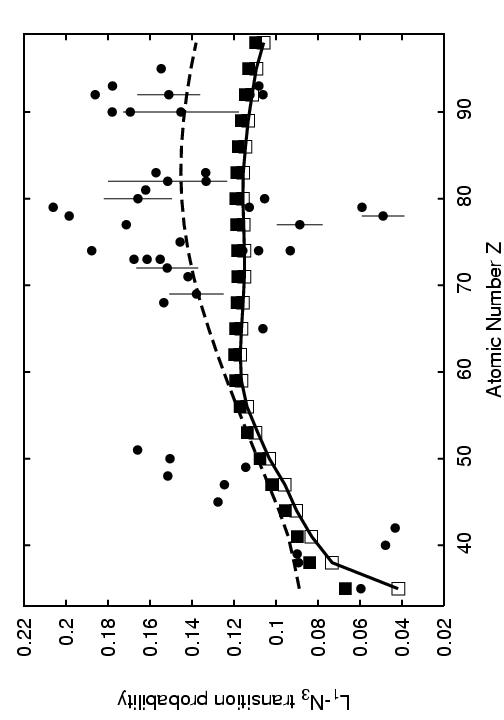}
\caption{L$_{1}$-N$_{3}$ transition probability versus Z:
theoretical calculations based on the Hartree-Slater \cite{sco1}
(white squares) and the Hartree-Fock \cite{sco2} (black squares)
potentials, EADL \cite{eadl} tabulations (solid line), experimental
data (black circles) and fit to them as in \cite{salem} (dashed line).}
\label{fig-l1n3}
\end{figure}

\begin{figure}[!t]
\centering
\includegraphics[width=2.5in,angle=-90]{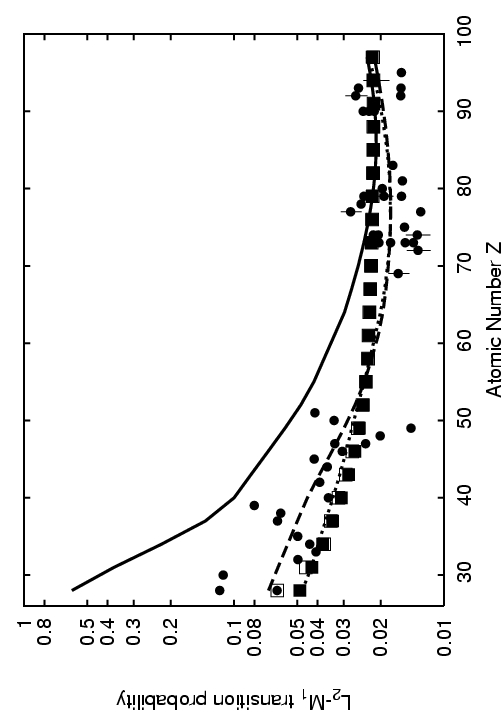}
\caption{L$_{2}$-M$_{1}$ transition probability versus Z:
theoretical calculations based on the Hartree-Slater \cite{sco1}
(white squares) and the Hartree-Fock \cite{sco2} (black squares)
potentials, EADL \cite{eadl} tabulations (solid line), experimental
data (black circles), fit to them as in \cite{salem} (dashed line), and
improved fit (dotted line).}
\label{fig-l2m1}
\end{figure}

\begin{figure}[!t]
\centering
\includegraphics[width=2.5in,angle=-90]{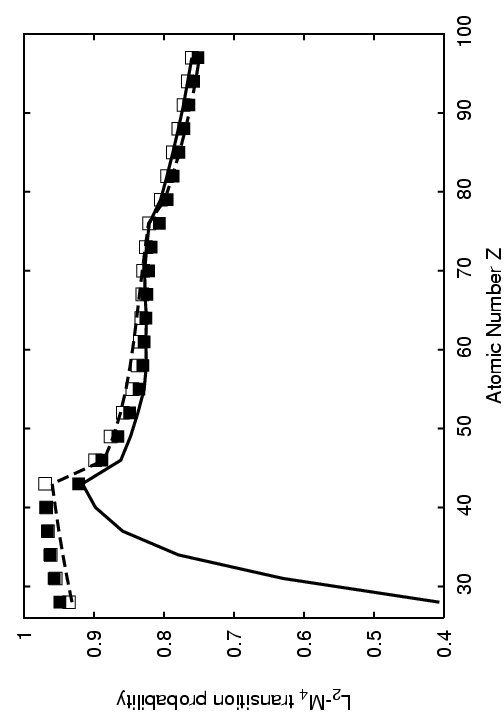}
\caption{L$_{2}$-M$_{4}$ transition probability versus Z:
theoretical calculations based on the Hartree-Slater \cite{sco1}
(white squares) and the Hartree-Fock \cite{sco2} (black squares)
potentials, EADL \cite{eadl} tabulations (solid line) and fit to experimental
data as in \cite{salem} (dashed line).}
\label{fig-l2m4}
\end{figure}

\begin{figure}[!t]
\centering
\includegraphics[width=2.5in,angle=-90]{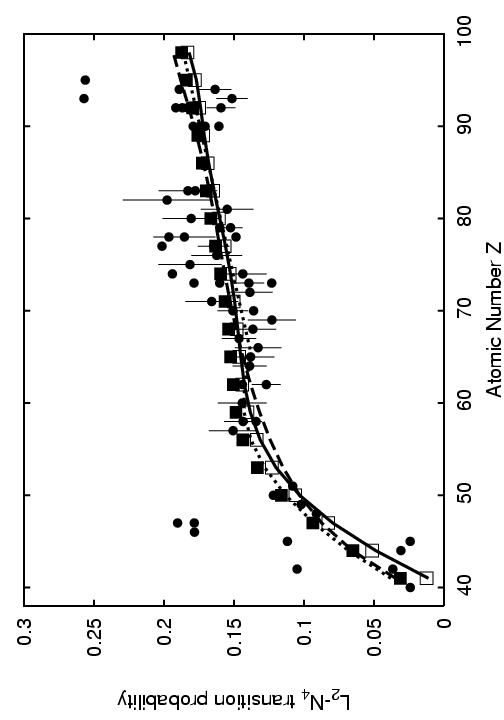}
\caption{L$_{2}$-N$_{4}$ transition probability versus Z:
theoretical calculations based on the Hartree-Slater \cite{sco1}
(white squares) and the Hartree-Fock \cite{sco2} (black squares)
potentials, EADL \cite{eadl} tabulations (solid line), experimental
data (black circles) fit to them as in \cite{salem} (dashed line), and
improved fit (dotted line).}
\label{fig-l2n4}
\end{figure}

\begin{figure}[!t]
\centering
\includegraphics[width=2.5in,angle=-90]{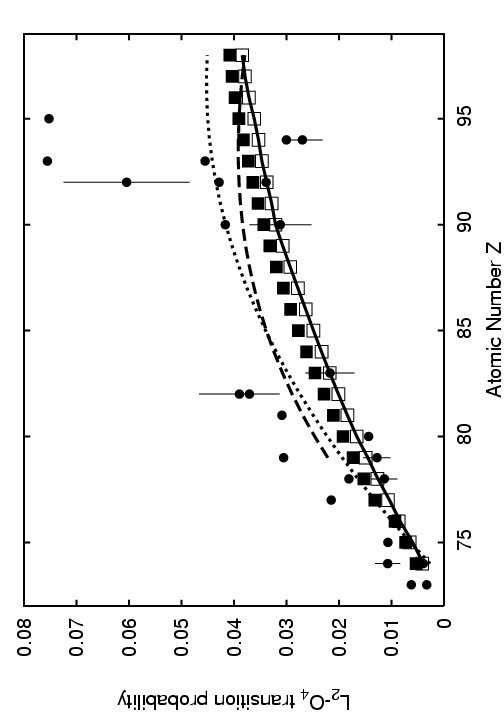}
\caption{L$_{2}$-O$_{4}$ transition probability versus Z:
theoretical calculations based on the Hartree-Slater \cite{sco1}
(white squares) and the Hartree-Fock \cite{sco2} (black squares)
potentials, EADL \cite{eadl} tabulations (solid line), experimental
data (black circles), fit to them as in \cite{salem} (dashed line), and
improved fit (dotted line).}
\label{fig-l2o4}
\end{figure}

\begin{figure}[!t]
\centering
\includegraphics[width=2.5in,angle=-90]{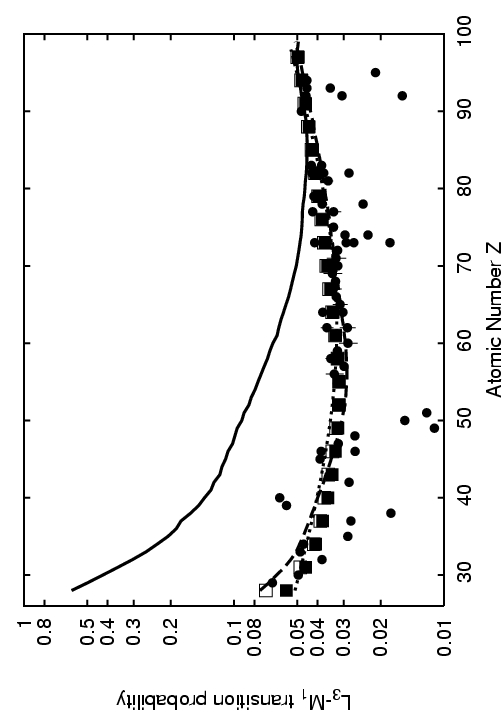}
\caption{L$_{3}$-M$_{1}$ transition probability versus Z:
theoretical calculations based on the Hartree-Slater \cite{sco1}
(white squares) and the Hartree-Fock \cite{sco2} (black squares)
potentials, EADL \cite{eadl} tabulations (solid line), experimental
data (black circles), fit to them as in \cite{salem} (dashed line), and
improved fit (dotted line).}
\label{fig-l3m1}
\end{figure}

\begin{figure}[!t]
\centering
\includegraphics[width=2.5in,angle=-90]{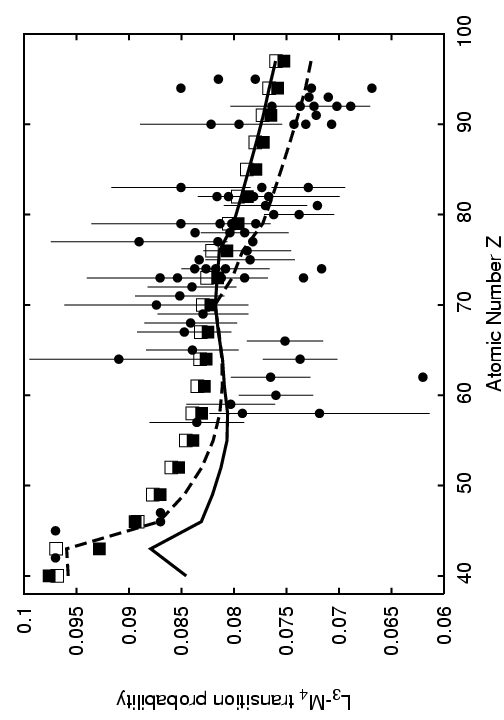}
\caption{L$_{3}$-M$_{4}$ transition probability versus Z:
theoretical calculations based on the Hartree-Slater \cite{sco1}
(white squares) and the Hartree-Fock \cite{sco2} (black squares)
potentials, EADL \cite{eadl} tabulations (solid line), experimental
data (black circles) and fit to them as in \cite{salem} (dashed line).}
\label{fig-l3m4}
\end{figure}

\begin{figure}[!t]
\centering
\includegraphics[width=2.5in,angle=-90]{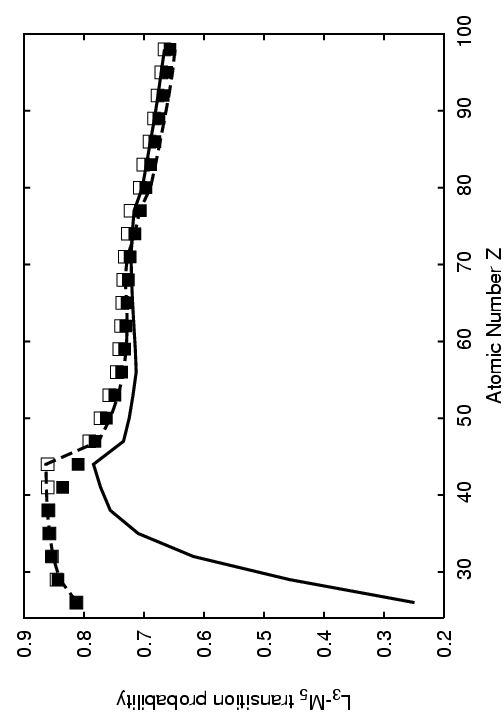}
\caption{L$_{3}$-M$_{5}$ transition probability versus Z:
theoretical calculations based on the Hartree-Slater \cite{sco1}
(white squares) and the Hartree-Fock \cite{sco2} (black squares)
potentials, EADL \cite{eadl} tabulations (solid line) and fit to experimental
data as in \cite{salem} (dashed line).}
\label{fig-l3m5}
\end{figure}

\begin{figure}[!t]
\centering
\includegraphics[width=2.5in,angle=-90]{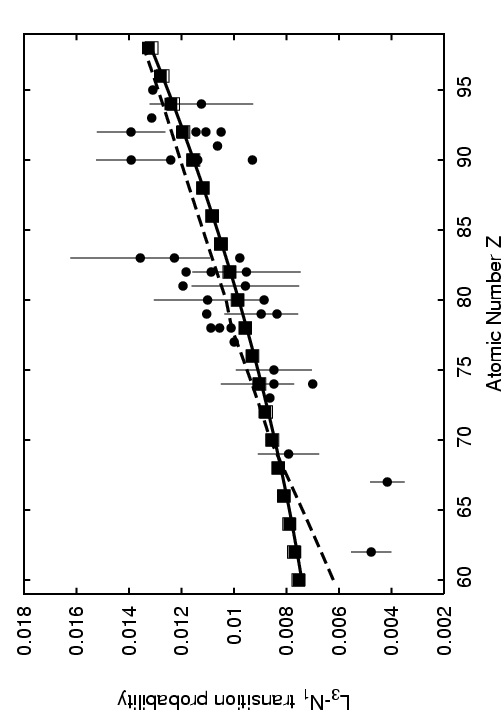}
\caption{L$_{3}$-N$_{1}$ transition probability versus Z:
theoretical calculations based on the Hartree-Slater \cite{sco1}
(white squares) and the Hartree-Fock \cite{sco2} (black squares)
potentials, EADL \cite{eadl} tabulations (solid line), experimental
data (black circles) and fit to them as in \cite{salem} (dashed line).}
\label{fig-l3n1}
\end{figure}

\begin{figure}[!t]
\centering
\includegraphics[width=2.5in,angle=-90]{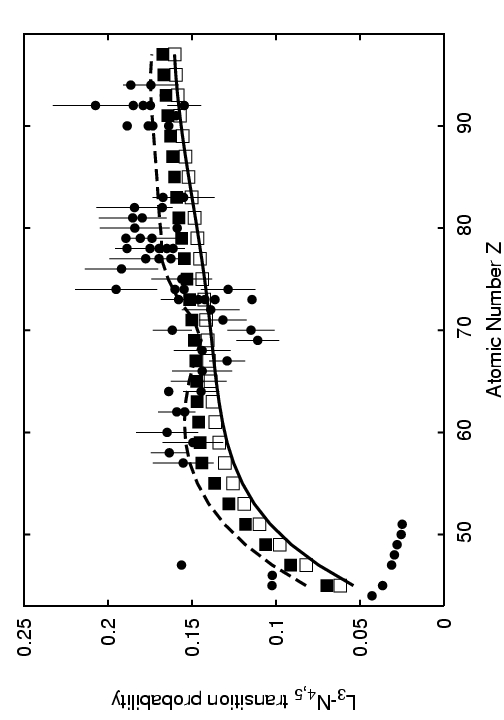}
\caption{L$_{3}$-N$_{4,5}$ transition probability versus Z:
theoretical calculations based on the Hartree-Slater \cite{sco1}
(white squares) and the Hartree-Fock \cite{sco2} (black squares)
potentials, EADL \cite{eadl} tabulations (solid line), experimental
data (black circles) and fit to them as in \cite{salem} (dashed line).}
\label{fig-l3n45}
\end{figure}

\begin{figure}[!t]
\centering
\includegraphics[width=2.5in,angle=-90]{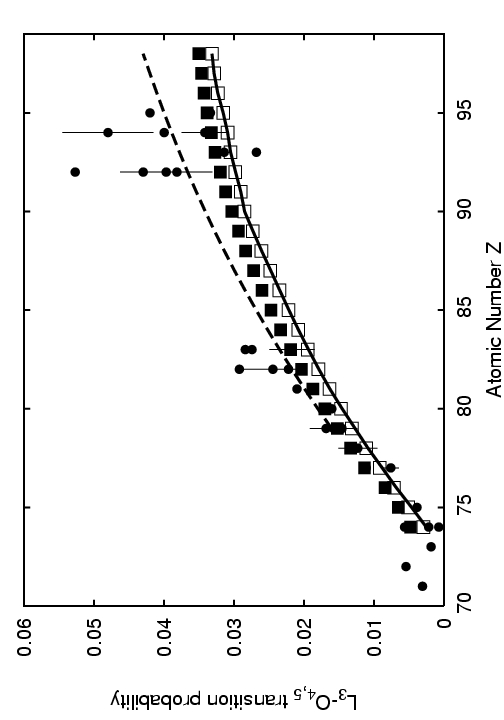}
\caption{L$_{3}$-O$_{4,5}$ transition probability versus Z:
theoretical calculations based on the Hartree-Slater \cite{sco1}
(white squares) and the Hartree-Fock \cite{sco2} (black squares)
potentials, EADL \cite{eadl} tabulations (solid line), experimental
data (black circles) and fit to them as in \cite{salem} (dashed line).}
\label{fig-l3o45}
\end{figure}

The transition probabilities for the K and L shells are shown in
Fig. \ref{fig-kl2} through \ref{fig-l3o45}.
The plots include the experimental data collected in \cite{salem}, the
Hartree-Slater and Hartree-Fock theoretical values, the corresponding
EADL values, the fits to the experimental data as in \cite{salem} and
\cite{elam}, and the improved fits mentioned in section \ref{sec_fit}.

The p-values resulting from the $\chi^{2}$ tests are listed in Table
\ref{tab_chi2} for each transition. 
The 
%values reported in italics correspond to the 
reference transitions in the probability ratios of \cite{salem}
appear in italic.

\begin{table}[!t]
\caption{p-values of the $\chi^{2}$ test comparing transition
probabilities from theoretical calculations 
against experimental data}
\label{tab_chi2}
\centering
%{\scriptsize
\begin{tabular}{|c|c|c|c|}
\hline
{\bf Transition}   	& {\bf Hartree}	&{\bf Hartree}  & {\bf EADL}           \\
                   	& {\bf Slater}  &{\bf Fock}     &                \\
\hline		                      		                          
K-L$_{2}$ 	   	& 0.025      	& 0.948   	& 0.024          \\
\textit{K-L$_{3}$} 	& 1       	& 1  	  	& 1     \\
K-M$_{2}$ 	   	& 0.953         & 0.407   	& 0.958          \\
K-M$_{3}$ 	   	& 1             & 1	   	& 1	           \\
K-M$_{4,5}$ 	   	& 0.110         & 0.849   	& 0.118          \\
K-N$_{2,3}$ 	   	& $<0.001$      & 0.717   	& $<0.001$       \\
K-N$_{4,5}$ 	   	& 0.033         & 0.192   	& 0.033          \\ 
\hline		        	               	       	  
L$_{1}$-M$_{2}$    	& 0.024         & 0.099   	& 0.024          \\
\textit{L$_{1}$-M$_{3}$} &0.158  	& 0.097   	& 0.384 \\
L$_{1}$-N$_{2}$    	& 0.186         & 0.283   	& 0.184           \\
L$_{1}$-N$_{3}$    	& 0.016         & 0.241   	& 0.016           \\ 
\hline		     	               	       		  	 
L$_{2}$-M$_{1}$    	& 0.001         & $<0.001$  	& $<0.001$        \\ 
\textit{L$_{2}$-M$_{4}$}    & 1       	& 1       	& 0      \\ 
L$_{2}$-N$_{4}$    	& 0.421         & 0.186   	& 0.398     	    \\ 
L$_{2}$-O$_{4}$    	& 0.006         & 0.110   	& 0.003     	    \\ 
\hline		     	                	       	  		
L$_{3}$-M$_{1}$    	& 0.289         & 0.455   	& $<0.001$            \\                                            
L$_{3}$-M$_{4}$    	& 0.721         & 0.880   	& 0.831           \\  
\textit{L$_{3}$-M$_{5}$}    & 1       	& 1        	& 1     \\  
L$_{3}$-N$_{1}$    	& 1             & 1	  	& 1	            \\  
L$_{3}$-N$_{4,5}$  	& 0.002         & 0.277   	& $<0.001$        \\  
L$_{3}$-O$_{4,5}$  	& 0.015         & 0.586   	& 0.004           \\  
\hline			  	        
\end{tabular}
\end{table}

Assuming a confidence level of 95\%, one can observe in Table \ref{tab_chi2}
that the $\chi^{2}$ test rejects the null hypothesis of equivalence 
of the distributions subject to the test for a larger number
of cases when comparing Hartree-Slater calculations to experimental
data, with respect to comparisons involving Hartee-Fock ones.
For transitions directly compared to experimental data (i.e. those listed
in the left column of Table \ref{tab_trans}), the null hypothesis
is rejected in 53\% of the test cases for the Hartree-Slater calculations,
while it is rejected in 6\% of the cases for the Hartree-Fock ones.
The rejection of the null hypothesis occurs in a slightly larger number of
the test cases (59\%) for EADL with respect to Hartree-Slater calculations.

An analysis based on contingency tables was performed to estimate
the statistical significance of the different accuracy observed 
with Hartree-Slater and EADL theoretical transition probabilities
with respect to the Hartree-Fock ones.
The contingency tables were based on the number of test cases
which pass or fail the $\chi^{2}$ test, assuming a 95\% confidence level 
for the rejection of the null hypothesis.
The transitions involving direct comparisons to experimental data
and the whole set of transitions were examined separately, to avoid
introducing a possible bias in the conclusions due to different
treatments of the data.
The results are summarized in Table \ref{tab_cont}.

\begin{table}[!t]
\caption{Contingency tables
comparing the accuracy of transition probability calculations}
\label{tab_cont}
\centering
\begin{tabular}{|c|c|c|}
\hline
\multicolumn{3}{|c|}{\textit{Transitions with direct experimental comparisons}} \\
\hline
{\bf $\chi^{2}$ test result} 	& {\bf Hartree-Slater} 	& {\bf Hartree-Fock}\\
%\hline
Pass 			& 8 			& 16 \\
Fail 			& 9 			& 1 \\ 
\hline
Fisher p-value 			& \multicolumn{2}{|c|}{0.007} \\ 
Yates $\chi^{2}$ p-value	& \multicolumn{2}{|c|}{0.008} \\ 
\hline
{\bf $\chi^{2}$ test result} 	& {\bf EADL} 		& {\bf Hartree-Fock}\\
%\hline
Pass 				& 7 			& 16 \\
Fail 				& 10 			& 1 \\ 
\hline
Fisher p-value 			& \multicolumn{2}{|c|}{0.002} \\ 
Yates $\chi^{2}$ p-value	& \multicolumn{2}{|c|}{0.003} \\ 
\hline
\multicolumn{3}{|c|}{\textit{All transitions}} \\
\hline
{\bf $\chi^{2}$ test result} 	& {\bf Hartree-Slater} 	& {\bf Hartree-Fock}\\
%\hline
Pass 	& 12 			& 20 \\
Fail 	& 9 			& 1 \\ 
\hline
Fisher p-value 			& \multicolumn{2}{|c|}{0.009} \\ 
Yates $\chi^{2}$ p-value	& \multicolumn{2}{|c|}{0.011} \\ 
\hline
{\bf $\chi^{2}$ test result} 	& {\bf EADL} 		& {\bf Hartree-Fock}\\
%\hline
Pass 	& 10 			& 20 \\
Fail 	& 11 			& 1 \\ 
\hline
Fisher p-value 			& \multicolumn{2}{|c|}{0.001} \\ 
Yates $\chi^{2}$ p-value	& \multicolumn{2}{|c|}{0.002} \\ 
\hline
\end{tabular}
\end{table}

% ----------------------------------------------------------------------------
\section{Comparative evaluations}

The statistical results reported in the previous section 
are the basis for comparative evaluations.

\subsection{Comparison of the accuracy of Hartree-Slater and
Hartree-Fock calculations}

The results of the statistical analysis in the previous section 
highlight a significant difference in the overall
accuracy of the Hartree-Slater and Hartree-Fock calculations
of radiative transition probabilities.
While the more refined nature of the Hartree-Fock approach has 
been known from a theoretical perspective, this study provides
a quantitative appraisal of the relative merits of the two
calculations with respect to a large experimental sample.

More precise experimental data over a large number of elements would
be needed to achieve firm conclusions of the relative accuracy of the
two methods for individual atomic transitions.

\subsection{Evaluation of EADL accuracy}

Some differences are observed in Table \ref{tab_chi2} regarding the
p-values related to the comparison of Hartree-Slater calculations and
EADL against the experimental references.
%while the same behaviour is expected, since both EADL 
%and the Hartree-Slater calculations considered in this study 
%nominally derive from \cite{sco0,sco1}.
Small differences in the test statistics could derive from the data
treatment described in the previous sections, which involves various
manipulations to normalize the data to common references;
however, large observed discrepancies should be ascribed to
other reasons, which cannot be elucidated based on EADL documentation
\cite{eadl}.

One can observe significant differences between the EADL values and the
Hartree-Slater calculations for the L$_{2}$-M$_{1}$ (Fig. \ref{fig-l2m1}),
L$_{2}$-M$_{4}$ (Fig. \ref{fig-l2m4}), L$_{3}$-M$_{1}$ (Fig. \ref{fig-l3m1}) and
L$_{3}$-M$_{5}$ (Fig. \ref{fig-l3m5}) transitions; some discrepancies are also
visible for L$_{3}$-M$_{4}$ (Fig. \ref{fig-l3m4}) and L$_{3}$-N$_{4,5}$
(Fig. \ref{fig-l3n45}).
Discrepancies against experimental data were observed for some of these
transitions in \cite{bonetto}.
Similar differences were observed \cite{relax_prob_nss}
between transition probabilities calculated by Geant4, which
uses EADL in its atomic relaxation package \cite{relax}, and the
fitted data in \cite{elam}.

Apart from these inconsistencies, the EADL content appears to reflect the
Hartree-Slater calculations in \cite{sco0,sco1}; therefore the comments about
the overall relative accuracy of Hartree-Slater calculations in the previous
section hold for EADL too.

The results of this study suggest that a revision of EADL 
would be desirable to include more accurate radiative transition 
probabilities based on Hartree-Fock calculations.

% ----------------------------------------------------------------------------
\section{Conclusion}
A systematic and quantitative validation of the existing theoretical
models for computing K and L shell fluorescence transition
probabilities was performed against an extensive collection of
experimental data.
The results, based on statistical methods, show that transition
probabilities derived from Hartree-Fock calculations better represent the
experimental measurements.

The EADL data library has been found not to represent the
state-of-the-art for what concerns radiative transition probabilities.
Based on the quantitative evidence obtained from this study, tabulations of
Hartree-Fock values can be recommended as a replacement for the current
radiative transition probabilities in EADL.
Such an update of EADL would contribute to improve the accuracy of the
Monte Carlo codes which use this data library for the simulation of
X-ray fluorescence.

% ----------------------------------------------------------------------------

% 1

% 2

% 3

% 4

%\clearpage
% 5

% 6

% 7

% ---- L1

% 8

%\clearpage
% 9

\clearpage

% 10

% 11

% ---- L2

% 12

% 13

% 14

%\clearpage
% 15

\clearpage
% ---- L3
% 16

%\clearpage
% 17

% 18

% 19

% 20

%\clearpage
% 21

\end{document}